\documentclass{sigcomm-alternate}

\usepackage{flushend} 

\usepackage{graphicx}
\usepackage{subfigure}

\usepackage{amsmath}
\usepackage{amsfonts}
\usepackage{mdwlist}
\usepackage{enumitem}

\usepackage{url}
\newcommand{\parheading}[1]{\medskip{} \noindent \textbf{#1}}

\usepackage{xcolor}
\newcommand{\cut}[1]{}

\title{A cost-benefit analysis of low latency via added utilization}

\numberofauthors{5}
\author{
\alignauthor
Ashish Vulimiri\\
\affaddr{UIUC}\\
\email{vulimir1@illinois.edu}\\
\alignauthor
P.\ Brighten Godfrey\\
\affaddr{UIUC}\\
\email{pbg@illinois.edu}\\
\alignauthor
Sri Varsha Gorge\\
\affaddr{Salesforce}\\
\email{gorge.srivarsha@gmail.com}\\
\and  
\alignauthor
Zitian Liu\\
\affaddr{Microsoft}\\
\email{liu236@illinois.edu}\\
\alignauthor
Scott Shenker\\
\affaddr{UC Berkeley and ICSI}\\
\email{shenker@icsi.berkeley.edu}
}

\date{}

\begin{document}
\maketitle

\begin{abstract}
Several recently proposed techniques achieve latency reduction by trading it off for some amount of additional bandwidth usage.  But how would one quantify whether the tradeoff is actually beneficial in a given system?  We develop an economic cost vs.\ benefit analysis for answering this question.  We use the analysis to derive a benchmark for wide-area client-server applications, and demonstrate how it can be applied to reason about a particular latency saving technique --- redundant DNS requests.
\end{abstract}

\section{Introduction}

Many techniques for improving latency in the Internet trade off some amount of extra bandwidth consumption for reduced latency.  Examples include DNS prefetching~\cite{ChromiumPrefetching}, redundant~\cite{Vulimiri2012,Vulimiri2013} and hedged~\cite{Dean2013} queries, and speculative TCP loss recovery mechanisms~\cite{Flach2013}.  But what is the true cost of the added overhead, and when is it outweighed by the latency reduction achieved?  In this brief note, we use an economic cost vs.\ benefit analysis to study these questions.  We consider the tradeoff between cost and benefit in a specific class of systems: wide-area client-server applications (such as web browsing, DNS queries, etc.) involving clients using consumer-level connectivity and service providers in the cloud.  The framework we develop here serves as a baseline; it can be refined or extended for other systems.

Our framework allows for various combinations of incentives at the servers and the clients.  In the common scenario where both servers and clients care exclusively about their own benefit, we show that any technique that saves more than $10$ ms of latency (in the mean or the tail, depending on the metric we are concerned with) for every kilobyte of extra traffic that it sends is useful, even with very pessimistic estimates for the additional cost induced at both clients and servers.  This is a conservative bound assuming the most expensive cost estimates we found; the threshold can be orders of magnitude lower in many realistic scenarios, such as when clients use DSL instead of cellular connectivity.

We develop a framework for comparing the cost of and benefit from latency-saving techniques (\S\ref{sec:framework}); use this framework to derive a benchmark for wide-area client-server applications (\S\ref{sec:analysis}); and demonstrate how the benchmark can be applied in practice via a case study (\S\ref{sec:case-study}).

\section{Framework}
\label{sec:framework}

Consider DNS prefetching, where web browsers pre-emptively initiate DNS lookups for links on a webpage to save latency if the user chooses to follow the link.  Prefetching adds overhead both to the client, which potentially sends DNS requests for more links than the user actually follows, and to the DNS infrastructure, which needs to service these additional requests.  The corresponding benefit is the latency reduction at the client when following a prefetched link, which also translates to an increase in expected ad revenue at the server~\cite{brutlag09}.  DNS prefetching affects several entities, including clients, servers, and network operators.

We account for the cost and benefit to all the stakeholders affected by any given latency-saving technique by comparing the following five quantitities:

\begin{itemize}[noitemsep]
\item $\ell$ (ms/KB): the average latency savings achieved by the technique, normalized by the volume of extra traffic it adds
\item $p_s$, $p_c$ (\$/KB): the average price of processing extra traffic at the servers and the clients
\item $v_s$, $v_c$ (\$/ms): the average value from latency improvement to the servers and the clients
\end{itemize}

We denote increased utilization in units of data transfer volume and measure added cost at the server and the client.   Note however that these calculated costs are a proxy for \emph{all} the costs (not just bandwidth) incurred by \emph{all} affected entities.  For instance, network operator costs are accounted for via the bandwidth costs ISPs charge servers and clients, and CDN costs are accounted for via the usage fees paid by servers.  One kilobyte of added client-side traffic in a web service has server-side costs including server utilization, energy, network operations staff, network usage fees, and so on.  In essence, we amortize all these diverse costs over units of client- and server-side traffic.

From the perspective of a selfish client, any latency-saving technique is useful as long as the benefit it adds outweights the cost to the client: that is, $\ell \times v_c \ge p_c$, or in other words
\[\ell \ge \frac{p_c}{v_c} \]

Similarly, a selfish server would need
\[\ell \ge \frac{p_s}{v_s} \]

\definecolor{darker-gray}{gray}{0.5}
\definecolor{lighter-gray}{gray}{0.7}
\newcommand{\deemphWeak}[1]{\textcolor{darker-gray}{#1}}
\newcommand{\deemphStrong}[1]{\textcolor{lighter-gray}{#1}}

\begin{table*}[ht!]
\begin{center}
{\footnotesize
\begin{tabular}{l|r@{\extracolsep{0pt}.}l|r@{\extracolsep{0pt}.}lr@{\extracolsep{0pt}.}l}
\hline
Service plan & \multicolumn{2}{c|}{Cost $c$} & \multicolumn{4}{c}{Break-even benefit $\ell$ (msec/KB), assuming...}\\
& \multicolumn{2}{c|}{(\$/GB)} & \multicolumn{2}{c}{server-side value} & \multicolumn{2}{c}{client-side value}\\
& \multicolumn{2}{c|}{} & \multicolumn{2}{c}{$v_s = \$1.54$/hr} & \multicolumn{2}{c}{$v_c = \$24.54$/hr}\\
\hline
\hline
\emph{Server-side plans} & \multicolumn{2}{c|}{$p_s$} & \multicolumn{2}{c}{~~~~~~~~$p_s/v_s$} & \multicolumn{2}{c}{\deemphWeak{$p_s/v_c$}~~~~~~~~}\\
~~~Amazon web services: ``Common Customer'' web app & 2&67 & ~~~~~~~~~~~~5&95 & ~~~~~~~~~~~~&\deemphWeak{37}\tabularnewline
~~~Amazon Route 53 (DNS) assuming 0.5KB/query & 1&40 & 3&12 & &\deemphWeak{20}\\
~~~Amazon CloudFront: U.S., 1 GB/mo, 1 KB/object & &91 & 2&03 & &\deemphWeak{13}\\
~~~Amazon EC2 and Microsoft Azure: bandwidth, Brazil & &25 & &56 & &\deemphWeak{035}\\
~~~NearlyFreeSpeech.net: web hosting & &25 & &56 & &\deemphWeak{035}\\
~~~Amazon CloudFront: U.S., 1 GB/mo, 10 KB/object & &20 & &45 & &\deemphWeak{028}\\
~~~Amazon EC2 and Microsoft Azure: bandwidth, US & &12 & &27 & &\deemphWeak{017}\\
~~~MaxCDN: based on ``starter'' plan overage fee & &08 & &18 & &\deemphWeak{011}\\
~~~DreamHost: cloud storage, object delivery & &075 & &17 & &\deemphWeak{010}\\
 & \multicolumn{2}{c|}{} & \multicolumn{2}{c}{} & \multicolumn{2}{c}{}\\
\emph{Client-side plans} & \multicolumn{2}{c|}{$p_c$} & \multicolumn{2}{c}{~~~~~~~~\deemphStrong{$p_c/v_s$}} & \multicolumn{2}{c}{$p_c/v_c$~~~~~~~~}\\
~~~AT\&T, low volume cell plan, based on overage fees & ~68&27 & \deemphStrong{152}&\deemphStrong{20} & 9&55\\
~~~AT\&T, high volume cell plan, based on overage fees & 15&00 & \deemphStrong{33}&\deemphStrong{44} & 2&10\\
~~~O$_{2}$ mobile broadband, based on 1GB$\to$2GB increment & 8&02 & \deemphStrong{17}&\deemphStrong{88} & 1&12\\
~~~AT\&T DSL & &20 & &\deemphStrong{45} & &028\\
\hline
\end{tabular}
}
\end{center}
\caption{Estimates of the cost of added utilization (in GB of data transfer), and resulting threshold benefit $\ell$ (in milliseconds saved per KB of added utilization) at which a technique becomes cost-effective. Based on providers' publicly advertised prices as of August 2014, excluding taxes and fees.}
\label{table:cost-benefit}
\end{table*}

We will require that both conditions be satisfied in the analysis in \S\ref{sec:analysis} -- that is, the benchmark we develop identifies latency-saving techniques that directly benefit both servers and clients.  Other combinations are possible.  For instance, a server might directly value both its own benefit as well as the improvement in user experience at the client, in which case we would need $\ell \ge \max \left\{ p_s / (v_s + v_c), p_c / v_c \right\}$.  The analysis can be modified to account for whatever incentives are necessary in any given application scenario.


\section{Analysis}
\label{sec:analysis}

\parheading{Cost estimates.}  To estimate server-side cost $p_s$, we use a range of advertised rates for cloud services which implement usage-based pricing, listed in the second column of Table~\ref{table:cost-benefit}. The most expansive (and expensive) of these is the first line, based on an Amazon Web Services sample customer profile of a web application.\footnote{\url{http://calculator.s3.amazonaws.com/calc5.html#key=a-simple-3-tier-web-app}}  The profile models a 3-tier auto-scalable web application, with a load balancer, two web servers, two app servers, a high-availability database server, 30 GB of storage, and other services, which utilizes $120$ GB/month of data transfer out of EC2 and $300$ GB/month out of CloudFront.  The resulting amortized cost of \$$2.67$ effectively models the cost (per transferred GB) of an average operation in this system, including the cost of all utilized services.\footnote{This is likely pessimistic since it includes, for example, the cost of increased storage which would not scale linearly with an increase in service operations.}  The other services listed in the table model the cost of more limited operations, such as DNS or bandwidth alone.

On the client side, we limit this investigation to clients in which the dominant cost of incrementally added utilization is due to network bandwidth.  Table~\ref{table:cost-benefit} lists costs $p_c$ based on several types of connectivity. For these calculations, we assume a user who has paid for basic connectivity already, and calculate the cost of bandwidth from overage charges. Client-side bandwidth costs can be substantially higher than server-side total costs in extreme (cellular) cases but are comparable or cheaper with DSL connectivity.

Of course, there are scenarios which the above range of application costs does not model. For example, a cellular client whose battery is nearly empty may value energy more than bandwidth.  But in a large class of situations, bandwidth is the most constrained resource on the client.


\parheading{Value estimates.} The value of time $v$ is more difficult to calculate, at both the client and server.

For the server, direct value may come from obtaining revenue (ads, sales).  We consider the case of Google.  A study by Google indicated that users experiencing an artifical $400$ ms added delay on each search performed $0.74$\% fewer searches after 4-6 weeks~\cite{brutlag09}.  Google's revenue per search has been estimated\footnote{Based on forecasts at \url{http://www.trefis.com}.} at \$$0.0231$; therefore, we can estimate a savings of $400$ milliseconds on a single search generates, on average, an additional $\$0.0231\cdot0.0074$ in revenue, or \$$1.54$ per hour of reduced latency.  As another estimate, a $500$ millisecond delay in the Bing search engine reduced revenue per user by $1.2$\%, or $4.3$\% with a $2$-second delay~\cite{souders09velocity}. Using the latter (smaller) figure, combined with an estimated\footnote{\url{http://www.trefis.com/company?hm=MSFT.trefis&driver=idMSFT.0817#}} revenue per Bing search of $\$0.0314$, we have a \$$2.43$ per hour value. We use the more pessimistic Google value of \$$1.54$/hr in our calculations.

On the client side, value may be obtained from a better or faster human experience.  Among all the components of our analysis, this value is the hardest to estimate: it may be highly application-specific, and may depend on mean or tail latency in ways best quantified by a human user study of quality of experience.  But as a first approximation, we assume the value of time is simply the US average earnings of \$24.54 per hour in August 2014~\cite{avgHourlyWage}, which implies $v_c \approx 6.82 \cdot 10^{-6} ~\$/\textrm{ms}$.

\parheading{Finding the threshold.}  We can now use our cost and value estimates to solve $\ell \geq p / v$ to obtain the break-even point, in terms of the necessary latency savings per kilobyte of additional traffic.

Table~\ref{table:cost-benefit} shows the break-even values of $\ell$ for various scenarios.  For example, the table indicates that a server replicating DNS traffic would obtain greater return in ad revenue than the cost of increased utilization with any latency-saving technique that saves more than $3.12$ milliseconds per KB of traffic that it adds.  The values are divided into four quadrants, one for each cost/benefit combination:

\newcommand{\subfigures}[1]{
\centering
\subfigure[Chrome, Emulab]{
\includegraphics[width=0.47\textwidth]{figures/emulab-#1}
}
\hspace{0.02\textwidth}
\subfigure[Firefox, residential DSL connection]{
\includegraphics[width=0.47\textwidth]{figures/residential-#1}
}
}

\begin{figure*}[p]
\subfigures{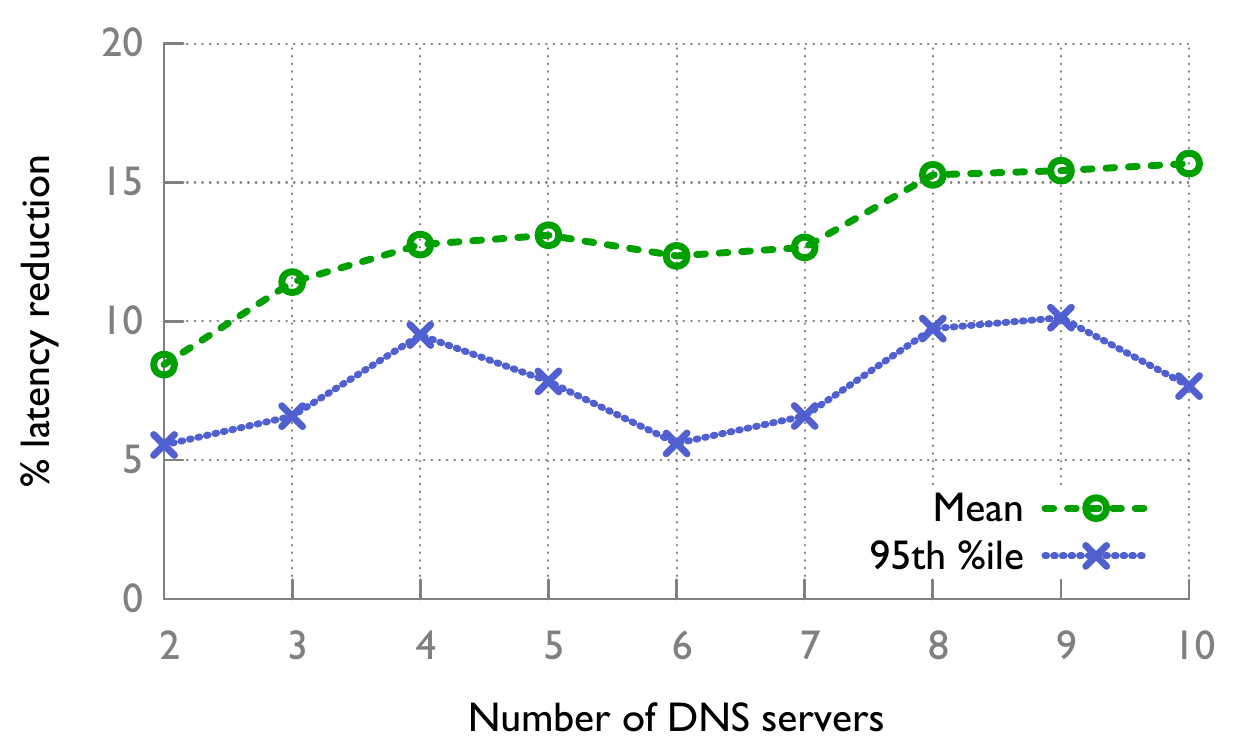}
\caption{Percentage improvement in page load times over unreplicated baseline}
\label{fig:relative}
\end{figure*}

\begin{figure*}[p]
\subfigures{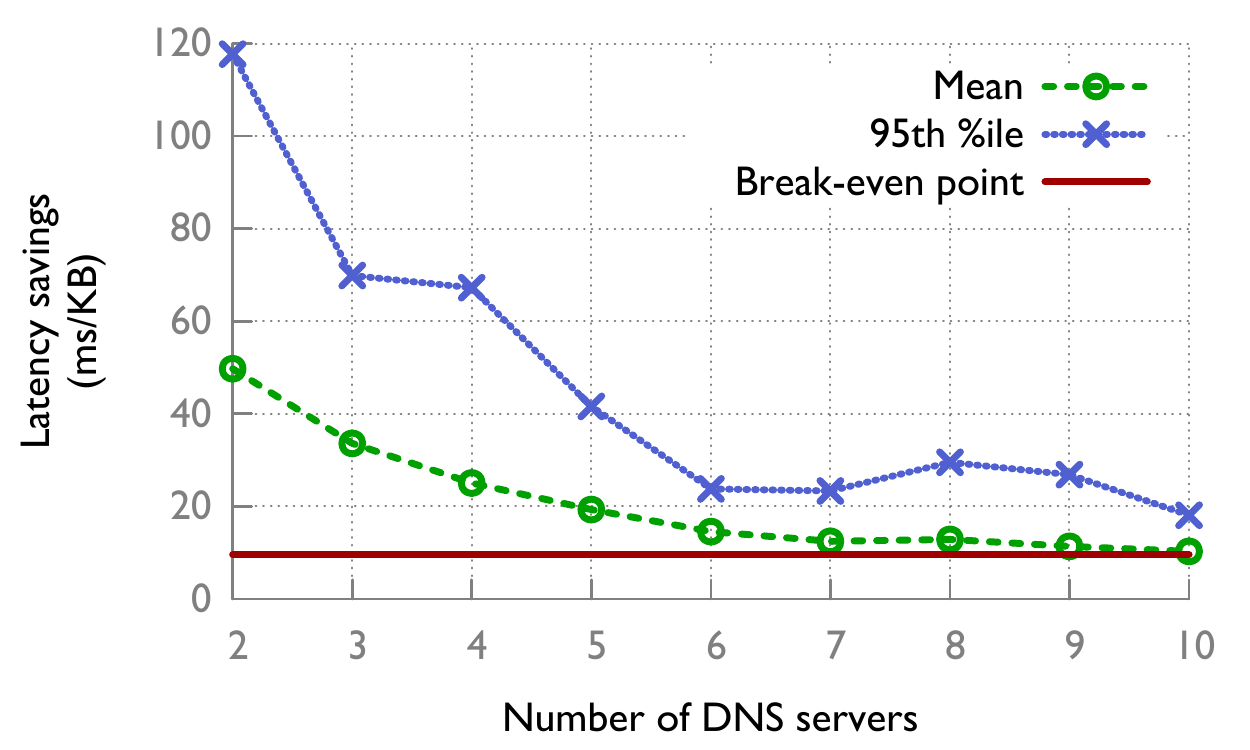}
\caption{Normalized latency savings at each level of replication}
\label{fig:ell-abs}
\end{figure*}

\begin{figure*}[p]
\subfigures{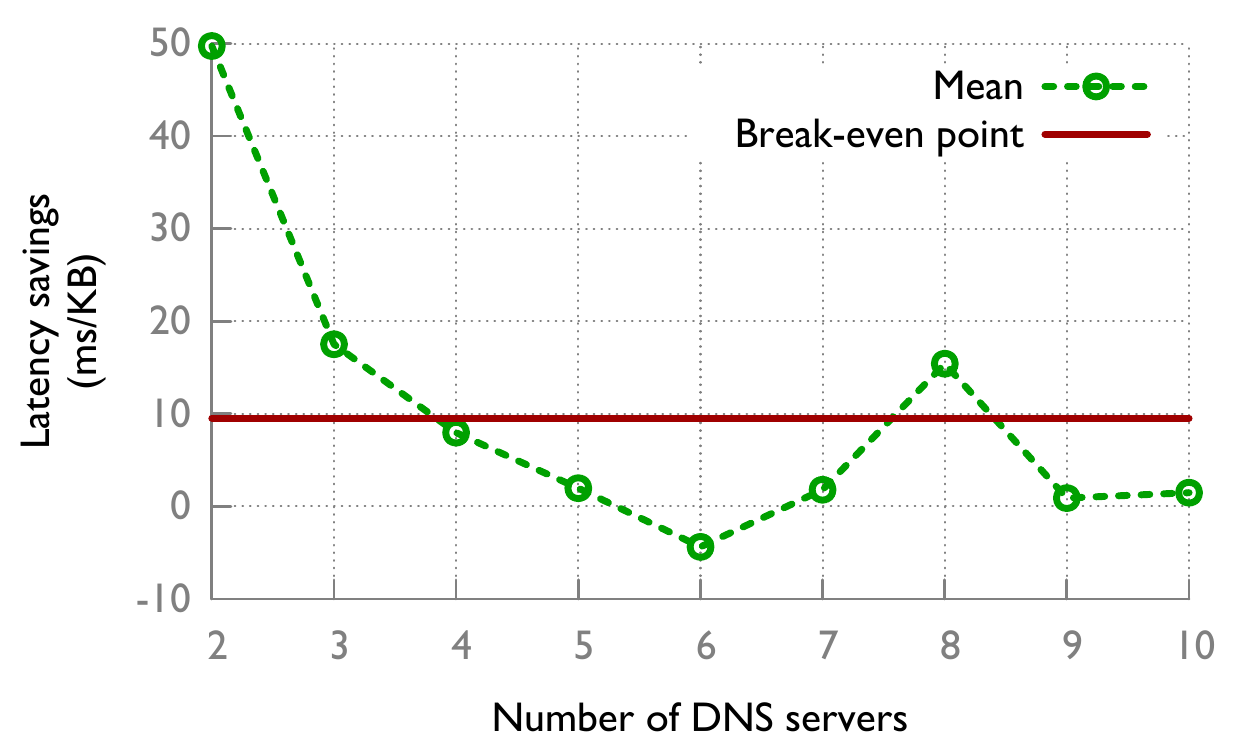}
\caption{Incremental latency savings from each additional DNS server}
\label{fig:ell-incr}
\end{figure*}

\begin{itemize*}
\item $p_s / v_s$ (upper-left quadrant): break-even $\ell$ for a server making a selfish decision.
\item $p_c / v_c$ (lower-right): a client making a selfish decision.
\item $p_s / v_c$ (upper-right): a server that directly values a client's quality of service.
\item $p_c / v_s$ (lower-left): a client that directly values the server's benefit.  This is unrealistic: a client would not typically value the server's ad revenue yet ignore its own benefit.
\end{itemize*}

Interestingly, excluding the last unrealistic scenario, both the server's and the client's worst-case break-even benefit occurs in a similar range of $6$-$10$ ms/KB; the client's higher costs are roughly balanced by its greater benefit.  This analysis suggests that a given technique may be cost-effective even in the most conservative cases as long as we can save more than $\approx$ 6-10 milliseconds (in the mean or tail, depending on the goal) for each kilobyte of added traffic.  Note that this is the worst-case value: the threshold can be orders of magnitude lower in many realistic scenarios.  For example, if clients use DSL (instead of cellular) connectivity and servers use an external web-host (instead of managing their own website on Amazon), the required latency savings threshold drops to 0.25 milliseconds per KB of added traffic.

%
%
%

\section{Case study: Redundant DNS\\requests}
\label{sec:case-study}

As an example, we now show how this benchmark can be applied to analyze a particular latency-saving technique, targeting DNS lookups: replicate DNS requests to one or more publicly accessible DNS servers in parallel, in addition to the default local ISP DNS server, and take the first reply that arrives~\cite{Vulimiri2012, Vulimiri2013}.

In previous work~\cite{Vulimiri2012,Vulimiri2013} we showed that this technique can achieve a significant reduction in DNS response times.  Replicating DNS requests to up to $10$ DNS servers in total we observed a $24 - 62\%$ lower request latency in the mean, median and tail than the unreplicated baseline.  The absolute improvement in all the metrics we tested was between $23$ and $761$ ms per KB of extra traffic added --- compared to the $10$ ms/KB benchmark we developed in \S\ref{sec:analysis}, this suggests that replicating DNS requests to $\le 10$ DNS servers is always cost-effective when considering raw DNS performance.

But protocol-level performance does not allow us to quantify client-side benefit.  Therefore, we now evaluate the \emph{appli\-cation-level} impact of replicating DNS requests, by quantifying the improvement in total web page load times when the technique is used.  We tested two deployments, Google Chrome running on an Emulab node, and Mozilla Firefox running on a laptop connected to the Internet via a residential DSL connection.  On both deployments we started with a list of $10$ DNS servers, the local ISP DNS server as well as $9$ publicly accessible DNS servers, and ran the following experiment: (1) Rank the list of DNS servers in order of their average DNS request latency.   (2)  Repeatedly pick a random website from Alexa.com's top-1000 list~\cite{alexa} and a random level of DNS replication $k \in [1, 10]$, and measure the time the browser takes to complete loading the website's homepage when every DNS request during the page load is replicated to the first $k$ DNS servers in the ranked list.  We set a $30$ second timeout and dropped all requests taking longer than $30$ seconds to complete (these were typically indicative of a failed script or a popup preventing the page from loading completely).

Figure~\ref{fig:relative} shows the percentage improvement in mean and 95th percentile page load times (compared to the unreplicated baseline) at various levels of replication.  In both deployments, we obtain a 6-15\% improvement in the two metrics, translating to an absolute improvement of 200-700 ms in the mean and 500-2300 ms in the 95th percentile.

Figure~\ref{fig:ell-abs} normalizes the observed improvement by the traffic overhead added and compares against the $10$ ms/KB benchmark from \S\ref{sec:analysis}.  The results show that replicating DNS queries to $10$ or more servers would be a net positive, both in the mean and the tail, in all the scenarios we analyzed in \S\ref{sec:analysis}.

Note that we observe diminishing returns: while the improvement generally increases with the level of redundancy, the \emph{incremental} improvement from each additional DNS server added to the system keeps decreasing.  At what point does adding servers cease to be cost-effective?  Figure~\ref{fig:ell-incr} answers this question by comparing the incremental improvement from each additional server against the $10$ ms/KB benchmark from \S\ref{sec:analysis}.  The results suggest that while replicating DNS requests to $2$ (perhaps $3$) DNS servers is better for mean latency than not replicating at all, $4$-way or higher replication is, at best, an economic net neutral.  We believe higher levels of replication may be appropriate in the tail, but at this point the data we have are too noisy to permit a similar analysis of tail latency; we are working on repeating these experiments at larger scales.

\section{Conclusion}

We proposed an analytical framework for evaluating techniques that improve latency by trading it off for additional bandwidth usage.  The analysis suggests a simple benchmark for client-server deployments in which both clients and servers care solely about their own benefit: any technique that improves latency by more than $10$ ms for each KB of extra traffic it adds is economically a net positive, even with very pessimistic estimates of the added costs at both servers and clients.  We showed how the analysis can be applied in practice by using it to identify the choices of parameters with which a particular latency-saving technique, DNS redundancy~\cite{Vulimiri2013,Vulimiri2012}, would prove beneficial.

\bibliographystyle{abbrv}
\bibliography{benchmark}

\end{document}